\documentclass[aps,prapplied,preprint,superscriptaddress]{revtex4-2}

\usepackage{epsfig,graphicx,times}
\usepackage{amstext}
\usepackage{amsmath}            
\usepackage{amssymb}            
\usepackage{graphicx}           
\usepackage{latexsym}
\usepackage{blindtext}
\usepackage{bm}
\usepackage{color}
\usepackage{multirow}
\usepackage[table,xcdraw]{xcolor}
\usepackage{lineno}

\makeatletter
\newenvironment{figurehere}
{\def\@captype{figure}}
{}
\makeatletter

\begin{document}

\title{Phase-controlled pathway interferences and switchable fast-slow light in a cavity-magnon polariton system}

\author{Jie Zhao}
\altaffiliation{These authors contributed equally to this work}
\affiliation{Hefei National Laboratory for Physical Sciences at the Microscale and Department of Modern Physics, University of Science and Technology of China, Hefei 230026, China}
\affiliation{CAS Key Laboratory of Microscale Magnetic Resonance, University of Science and Technology of China, Hefei 230026, China}
\affiliation{Synergetic Innovation Center of Quantum Information and Quantum Physics, University of Science and Technology of China, Hefei 230026, China}
\affiliation{National Laboratory of Solid State Microstructures, School of Physics, Nanjing University, Nanjing 210093, China}

\author{Longhao Wu}
\altaffiliation{These authors contributed equally to this work}
\affiliation{Hefei National Laboratory for Physical Sciences at the Microscale and Department of Modern Physics, University of Science and Technology of China, Hefei 230026, China}
\affiliation{CAS Key Laboratory of Microscale Magnetic Resonance, University of Science and Technology of China, Hefei 230026, China}
\affiliation{Synergetic Innovation Center of Quantum Information and Quantum Physics, University of Science and Technology of China, Hefei 230026, China}

\author{Tiefu Li}
\affiliation{Institute of Microelectronics, Tsinghua University, Beijing 100084, China}
\affiliation{Quantum states of matter, Beijing Academy of Quantum Information Sciences, Beijing 100193, China}

\author{Yu-xi Liu}
\affiliation{Institute of Microelectronics, Tsinghua University, Beijing 100084, China}

\author{Franco Nori}
\affiliation{Theoretical Quantum Physics Laboratory, RIKEN, Saitama, 351-0198, Japan}
\affiliation{Department of Physics, The University of Michigan, Ann Arbor, Michigan 48109-1040, USA}

\author{Yulong Liu}
\email[]{liuyl@baqis.ac.cn}
\affiliation{Beijing Academy of Quantum Information Sciences, Beijing 100193, China}
\affiliation{Department of Applied Physics, Aalto University, P.O. Box 15100, FI-00076 Aalto, Finland}

\author{Jiangfeng Du}
\email[]{djf@ustc.edu.cn}
\affiliation{Hefei National Laboratory for Physical Sciences at the Microscale and Department of Modern Physics, University of Science and Technology of China, Hefei 230026, China}
\affiliation{CAS Key Laboratory of Microscale Magnetic Resonance, University of Science and Technology of China, Hefei 230026, China}
\affiliation{Synergetic Innovation Center of Quantum Information and Quantum Physics, University of Science and Technology of China, Hefei 230026, China}

\date{\today}

\begin{abstract}
We study the phase controlled transmission properties in a compound system consisting of a 3D copper cavity and an yttrium iron garnet (YIG) sphere. By tuning the relative phase of the magnon pumping and cavity probe tones, constructive and destructive interferences occur periodically, which strongly modify both the cavity field transmission spectra and the group delay of light. Moreover, the tunable amplitude ratio between pump-probe tones allows us to further improve the signal absorption or amplification, accompanied by either significantly enhanced optical advance or delay. Both the phase and amplitude-ratio can be used to realize in-situ tunable and switchable fast-slow light. The tunable phase and amplitude-ratio lead to the zero reflection of the transmitted light and an abrupt fast-slow light transition. Our results confirm that direct magnon pumping through the coupling loops provides a versatile route to achieve controllable signal transmission, storage, and communication, which can be further expanded to the quantum regime, realizing coherent-state processing or quantum-limited precise measurements.
\end{abstract}
\maketitle

\section{Introduction}
Interference, due to superposed waves, plays a considerable role in explaining many classical and quantum physical phenomena. Based on the phase-difference-induced interference patterns, ultraprecise interferometers have been created, impacting the development of modern physics and industry~\cite{born1999principles}. In addition to the phases, waves or particles propagating through different pathways can also introduce interference patterns. Among various types of multiple-path-induced interference, the Fano resonance~\cite{miroshnichenko2010fano} and its typical manifestations, the electromagnetically induced transparency (EIT) and electromagnetically induced absorption (EIABS)~\cite{fleischhauer2005electromagnetically, anisimov2011objectively}, are the most well-known ones. The Fano resonance and EIT-like (or EIABS-like) line shapes are not only experimentally observed in quantum systems but also in various classical harmonic-resonator systems. Quantum examples include quantum dots~\cite{kroner2008nonlinear}, quantum wells~\cite{golde2009fano}, superconducting qubits~\cite{zhou2008controllable, ian2010tunable, Liu2016Method, long2018electromagnetically}, as well as Bose-Einstein condensates~\cite{akram2017control}. Classical examples~\cite{garrido2002classical} include coupled optical cavities~\cite{limonov2017fano, liu2017electromagnetically, peng2014and, ozdemir2019parity}, terahertz resonators~\cite{zhao2019terahertz, yahiaoui2018electromagnetically}, microwave resonators~\cite{johansson2014optomechanical, eichler2018realizing}, mechanical resonators~\cite{xu2016topological, liu2010acoustic}, optomechanical systems~\cite{aspelmeyer2014cavity}. However, whether in quantum or in classical systems, the Fano resonance, EIT- or EIABS-like spectra are normally experimentally realized separately. The switchable electromagnetically induced transparency and absorption, as well as fast and slow light, have been proposed using dressed superconducting qubits~\cite{ian2010tunable}, hybrid optomechanical system~\cite{akram2015tunable,li2016transparency}, dark-mode breaking~\cite{lai2020tunable,lake2020two,kuzyk2017controlling}, and so on. Particularly, there appears growing interest to control the EIT and EIABS by introducing exceptional points~\cite{liu2017controllable, wang2019mechanical,lu2018opto}. Photon stops~\cite{goldzak2018light,yang2020unconventional}, chiral EIT~\cite{wang2020electromagnetically}, and infinite slow light~\cite{yang2020unconventional} have recently been realized around exceptional points. Motivated by their potential applications in rapid transitions between fast and slow light, which facilitate coherent state storage and retrieval, it is highly desirable to have experimental realizations of \textit{in situ} tunable and switchable absorption, transparency, and even amplification.

Meanwhile, cavity magnon polaritons in an yttrium-iron-garnet (YIG) sphere-cavity coupled system has attracted much attention due to its strong~\cite{goryachev2014high, tabuchi2014hybridizing, zhang2015cavity, zhang2017observation, yao2017cooperative, bai2017cavity, kaur2016voltage, boventer2020control} and even ultrastrong couplings~\cite{kostylev2016superstrong, flower2019experimental, bourhill2016ultrahigh}. The compatibility and scalability with microwave and optical light enable magnons to be a versatile interface for different quantum devices~\cite{lachance2019hybrid, hisatomi2016bidirectional, zhang2016optomagnonic, kusminskiy2016coupled, kusminskiy2019cavity, liu2019phase}. At low temperatures, strong coupling between magnons, superconducting resonators and qubits have been demonstrated~\cite{tabuchi2015coherent, tabuchi2016quantum, morris2017strong, li2019strong, hou2019strong}. Subsequently, the EIT-like magnon-induced transparency (MIT) or the EIABS-like magnon-induced absorption (MIABS) of the transmitted cavity field were observed for different external coupling conditions~\cite{zhang2014strongly}. The underlying mechanism is attributed to interferences between two transition pathways, i.e., the direct cavity pathway and the cavity-magnon-cavity pathway, to transmit the probe field.

In addition to the coupling strength~\cite{zhang2014strongly} and frequency detuning~\cite{xu2019cavity,yang2019control,harder2018level} between coupled modes, phases play a vital role in wave interference control. We thus focus on the controllability of pathway interferences through the phase difference between the cavity-probe tone and the magnon-pump tone, which is introduced by the coupling loops' technology~\cite{wang2016magnon,wang2018bistability,zhao2020observation,boventer2019steering}. The direct magnon pump is becoming useful in realizing the light-wave interface~\cite{lachance2019hybrid, hisatomi2016bidirectional,zhang2016optomagnonic}, enhancing the Kerr nonlinearity~\cite{kong2019magnon,zhang2019theory,huai2019enhanced}, and has also been adopted to observe the magnetostriction-induced quantum entanglement~\cite{zhang2016cavity,li2018magnon,li2019entangling,li2019squeezed,yu2020magnetostrictively}, among other applications.

Together with the cavity-probe tone, a magnon-pump tone introduces a controllable relative phase to the system, and thus the path interference can be real-time controlled. Changing the two-tone phase difference, we can \textit{switch the cavity-probe spectra from the original magnon-induced transparency instantly to the magnon-induced absorption, or even the Fano line shape}. Furthermore, the tunable pump-probe amplitude ratio allows us to further improve the signal absorption, transparency, or amplification, accompanied by a significant enhancement by nearly \textit{2 orders of magnitude} of the optical advance or delay time compared to the case without magnon pump~\cite{zhang2014strongly}. In particular, the tunable phase and amplitude ratio also lead to the zero reflection of the transmitted light, which is accompanied by an abrupt transition of delay time. Our results confirm that direct magnon pumping provides a versatile route to control signal transmission, storage, and communication, and can be further expanded to coherent state processing in the quantum regime.

\begin{figurehere}
\begin{center}
\includegraphics[width=0.66\linewidth]{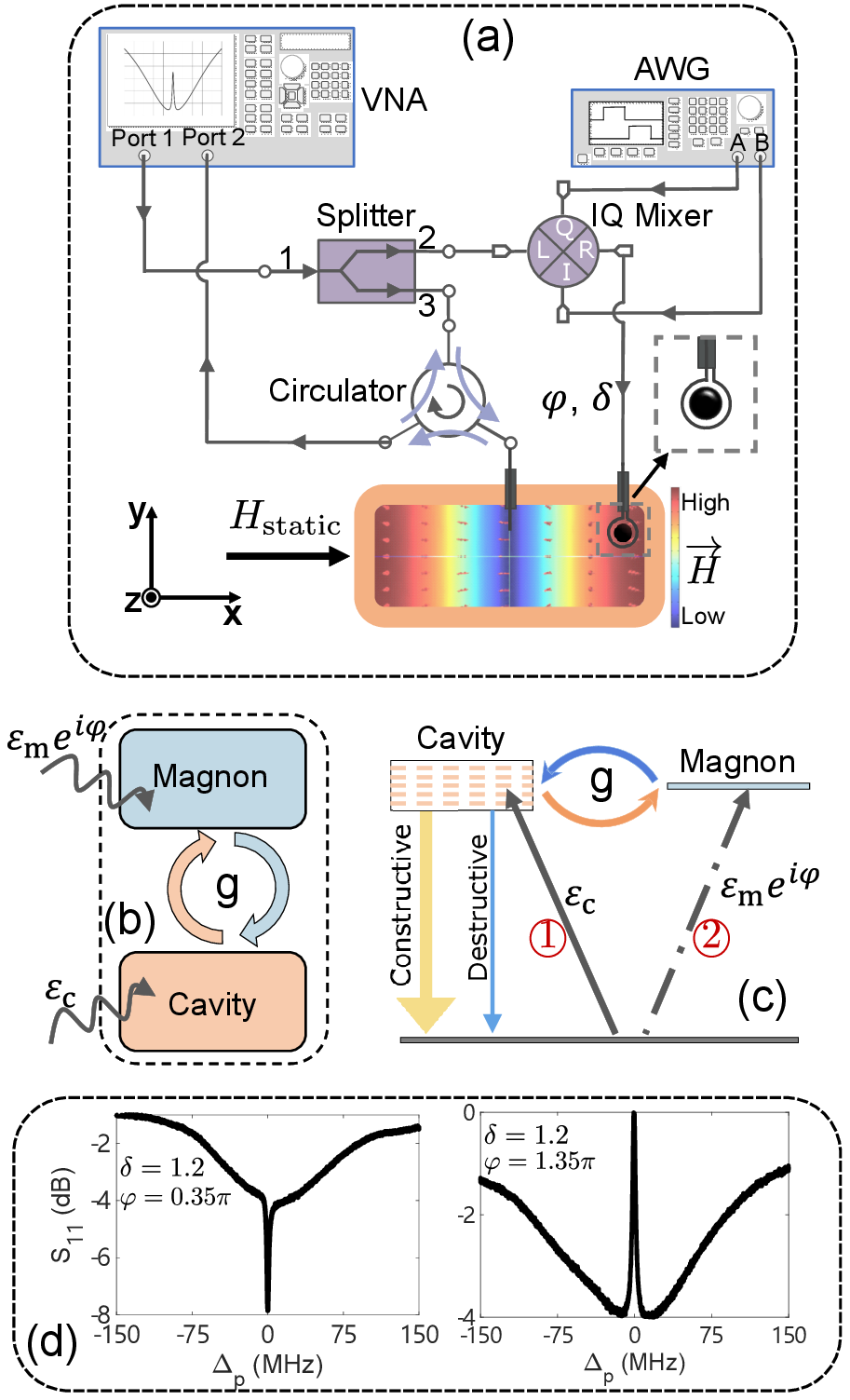}
\caption{\label{fig1} Measurement setup and phase-induced interference mechanism diagrams. (a) The system consisting of a three-dimensional (3D) copper cavity and a YIG sphere, which is coherently pumped by the coupling loops shown as a black coil surrounding the YIG sphere. The red arrows and colors indicate the magnetic field directions and amplitudes of the TE$_{101}$ mode distribution, respectively. The YIG sphere is placed at the area with maximum magnetic field distribution inside a 3D copper cavity box to obtain a strong cavity-magnon coupling. A small hole at the cavity sidewall is assembled with a standard SubMiniature version A connector (SMA connector), allowing us to do the reflection measurement $\rm S_{11}$ of the probe field, i.e., such a SMA connector works as both the signal input and readout port. A beam of coherent microwave comes out from port 1 of the vector network analyzer (VNA) and splits into two beams, working as the magnon-pump tone and the cavity-probe tone. Here, we use an in-phase and quadrature mixer (I-Q mixer) and an arbitrary waveform generator (AWG) to control and tune the phase difference $\varphi$ and pump-probe amplitude ratio $\delta=\varepsilon_{\mathrm{m}}/\varepsilon_{\mathrm{c}}$ between the pump and probe tones. The interfering results are extracted by the circulator and finally transferred to port 2 of the VNA. (b) Diagram showing the relative phase between the magnon pump and cavity probe in the cavity-magnon coupled system. (c) The corresponding energy-level diagram. Two transition pathways to the higher energy level: \textcircled{1} probe-tone-induced direct excitation, and \textcircled{2} pump-tone excites magnons and then coherently transfers there to cavity photons. (d) Measurements of the reflection spectra $\rm S_{11}$ versus the detuning $\Delta_{\mathrm{p}}=\omega_{\mathrm{c}}-\omega_{\mathrm{p}}=\omega_{\mathrm{m}}-\omega_{\mathrm{p}}$. The relative phase difference between pump and probe tones can be developed to realize an \textit{in situ} switchable constructive and destructive interference, presented as MIABS with $\varphi=0.35\pi,\delta=1.2$ and MIT with $\varphi=1.35\pi,\delta=1.2$.}
\end{center}
\end{figurehere}

\section{Experimental setup}
As shown in Fig.~\ref{fig1}(a), our system consists of a 3D copper (Cu) cavity with an inner dimension of $40\times20\times8$ $\textrm{mm}^{3}$ and an YIG sphere with a 0.3~mm diameter. A static magnetic field $H_\textrm{static}$ applied in the $x\mbox{-}y$ plane tunes the magnon frequency. The simulated cavity-mode magnetic field distribution is shown at the bottom of Fig.~\ref{fig1}(a), where the arrows and colors indicate the cavity mode magnetic field directions and amplitudes. The YIG sphere is placed near the magnetic field antinode of the cavity $\rm TE_{101}$ mode. The magnetic components (along the $z$ axis) of the microwave field at this antinode is perpendicular to the static magnetic bias field.

Here, we are only interested in the low excited states of the Kittel mode, in which all the spins precess in phase. Under the Holstein-Primakoff transformation, such collective spin mode can be simplified to a harmonic resonator, which introduces the magnon mode. In our setup, the cavity mode couples to the magnon mode with coupling strength $g=7.6\ \rm MHz$, which is larger than the magnon decay rate $\kappa_{\mathrm{m}}=1.2\ \rm MHz$, but smaller than the cavity decay rate $\kappa_{\mathrm{c}}=113.9\ \rm MHz$.

In our experiment, a beam of coherent microwave is emitted from port 1 of a VNA and then divided through a splitter into two beams, one of which is used to probe the cavity (probe tone) and another beam is used to pump the magnon (pump tone) by incorporating the coupling loop technique, which is schematically shown in the dashed rectangle of Fig.~\ref{fig1}(a). The probe tone is injected into the cavity through antenna 1, which induces the cavity external decay rate $\kappa_{\mathrm{c1}}=21.8\ \rm MHz$. The pump tone is injected through antenna 2, which introduces the magnon external decay rate $\kappa_{\mathrm{m1}}=0.6\ \rm MHz$. Note that the phase $\varphi _{\mathrm{c}}=0$ and amplitude $\varepsilon_{\mathrm{c}}$ of the probe tone are fixed (i.e., working as a reference), and the phase $\varphi$ and amplitude $\varepsilon _{\mathrm{m}}$ of the magnon-pump tone are tunable and controlled by an arbitrary wave generator with an in-phase and quadrature mixer (I-Q mixer).

\section{Model}
By considering the cavity-magnon coupling, as well as the pump and probe tones [model in Fig.~\ref{fig1}(b)], the system Hamiltonian becomes

\begin{eqnarray} \label{Ham}
H &=& \omega _{\mathrm{c}}a^{\dag }a + \omega _{\mathrm{m}}m^{\dag}m + g(a^{\dag }m+m^{\dag }a) \notag \\
&&+i\sqrt{2\eta _{\mathrm{c}}\kappa _{\mathrm{c}}}\; \varepsilon _{\mathrm{c}}\left( a^{\dag }e^{-i\omega _{\mathrm{p}}t}-ae^{i\omega _{\mathrm{p}}t}\right)   \notag \\
&&+i\sqrt{2\eta _{\mathrm{m}}\kappa _{\mathrm{m}}}\; \varepsilon _{\mathrm{m}}\left( m^{\dag }e^{-i\omega _{\mathrm{p}}t-i\varphi}-me^{i\omega _{\mathrm{p}}t+i\varphi}\right) .
\end{eqnarray}

\noindent Here, $a^{\dag }$ ($a$) and $m^{\dag }$ ($m$) are the creation (annihilation) operators for the microwave photon and the magnon at frequencies $\omega_{\mathrm{c}}$ and $\omega _{\mathrm{m}}$, respectively, and we choose units with $\hbar=1$. The magnon frequency $\omega _{\mathrm{m}}$ linearly depends on the static bias field $H_\textrm{static}$ and is tunable within the range of a few hundred MHz to about 45 GHz; $\varepsilon_{\mathrm{c}}$ ($\varepsilon_{\mathrm{m}}$) is the microwave amplitude applied to drive the cavity (magnon). Here, we introduce the coupling parameter

\begin{eqnarray}
\eta_{\mathrm{c}} = \kappa_{\mathrm{c1}}/\kappa_{\mathrm{c}}, \\
\eta_{\mathrm{m}} = \kappa_{\mathrm{m1}}/\kappa_{\mathrm{m}}
\end{eqnarray}

\noindent to classify the working regime of the cavity (the magnon). The parameter $\eta_{\mathrm{c}}$ ($\eta_{\mathrm{m}}$) classifies three working regimes for the cavity (magnon) into three types: overcoupling regime for $\eta_{\mathrm{c}}\ \left(\eta_{\mathrm{m}}\right) > 1/2$; critical-coupling regime for $\eta_{\mathrm{c}}\ \left(\eta_{\mathrm{m}}\right)=1/2$; and undercoupling regime for $\eta_{\mathrm{c}}\ \left(\eta_{\mathrm{m}}\right)<1/2$. In our experiment, the cavity works in the undercoupling regime ($\eta_{\mathrm{c}}<1/2$) and the magnon works in the critical coupling regime ($\eta_{\mathrm{m}}=1/2$).

Experimentally, the reflection signal from the cavity is circulated and then transferred to port 2 of the VNA to carry out the spectroscopic measurement, which corresponds to the steady-state solution of the Hamiltonian Eq.~(\ref{Ham}). The transmission coefficient $t_{\mathrm{p}}$ of the probe field is defined as the ratio of the output-field amplitude $\varepsilon_{\mathrm{out}}$ to the input-field amplitude $\varepsilon_{\mathrm{c}}$ at the probe frequency $\omega_{\mathrm{p}}$: $t_{\mathrm{p}}=\varepsilon_{\mathrm{out}}/\varepsilon_{\mathrm{c}}$. With the input-output boundary condition,

\begin{equation}
\varepsilon _{\mathrm{out}}=\varepsilon _{\mathrm{c}} - \sqrt{2\eta _{\mathrm{c}}\kappa _{\mathrm{c}}}\left\langle a\right\rangle,
\end{equation}

\noindent we can solve the transmission coefficient $t_{\mathrm{p}}$ of the probe field as~\cite{suppl}
\vspace*{-5mm}

\begin{equation} \label{trancoff}
t_{\mathrm{p}} = t_{\mathrm{probe}}+t_{\mathrm{pump}},
\end{equation}

\noindent with
\vspace*{-10mm}

\begin{eqnarray}
t_{\mathrm{probe}} &=& 1 - \frac{2\eta _{\mathrm{c}}\kappa _{\mathrm{c}}\left(i\Delta _{\mathrm{p}}+\kappa _{\mathrm{m}}\right)}{\left( i\Delta _{\mathrm{p}}+\kappa _{\mathrm{c}}\right) \left( i\Delta _{\mathrm{p}}+\kappa_{\mathrm{m}}\right) + g^{2}}, \label{tprobe} \\
t_{\mathrm{pump}} &=&\frac{ig\sqrt{2\eta _{\mathrm{c}}\kappa _{\mathrm{c}}}\sqrt{2\eta _{\mathrm{m}}\kappa _{\mathrm{m}}}\delta e^{-i\varphi }}{\left(i\Delta _{\mathrm{p}}+\kappa _{\mathrm{c}}\right) \left( i\Delta _{\mathrm{p}}+\kappa _{\mathrm{m}}\right) +g^{2}} . \label{tpump}
\end{eqnarray}

\noindent Here $\Delta_{\mathrm{p}}$ is the detuning between the probe frequency $\omega_{\mathrm{p}}$ and either the cavity resonant frequency $\omega_{\mathrm{c}}$ or the magnon frequency $\omega_{\mathrm{m}}$. In our experiment, the cavity is resonant with the cavity, i.e.,
\vspace*{-5mm}

\begin{equation}
\Delta_{\mathrm{p}}=\omega_{\mathrm{c}}-\omega_{\mathrm{p}}=\omega_{\mathrm{m}}-\omega_{\mathrm{p}};
\end{equation}

\noindent and
\vspace*{-10mm}

\begin{equation}
\delta =\varepsilon _{\mathrm{m}}/\varepsilon _{\mathrm{c}}
\end{equation}

\noindent is the pump-probe amplitude ratio. Equation~(\ref{trancoff}) clearly shows that the transmission coefficient can be divided into two parts:

\begin{enumerate}
\item $t_{\mathrm{probe}}$ in Eq.~(\ref{tprobe}), the contribution from the cavity-probe tone, represents the traditional pathway-induced interference;
\item $t_{\mathrm{pump}}$ in Eq.~(\ref{tpump}), the contribution from the magnon-pump field, affects the interference and modifies the transmission of the probe field.
\end{enumerate}

\noindent As shown in Fig.~\ref{fig1}(c), there exist two transition pathways for the cavity: the probe-tone-induced direct excitation, and the photons transferred from magnon excitations. When the cavity decay rate (analog to broadband of states) is much larger than the magnon decay rate (analog to a narrow discrete quantum state in other quantum systems), Fano interference happens and has been successfully used to explain the MIT and MIABS phenomenon in cavity magnon-polariton systems~\cite{zhang2014strongly}. Besides pathway-induced interference, the steered phase $\varphi$ of the wave provides another useful way to generate and especially control the interferences, as shown in Fig.~\ref{fig1}(d).

We emphasize that in this paper we focus on \textit{how the phase difference $\varphi$ and pump-probe ratio $\delta=\varepsilon _{\mathrm{m}}/\varepsilon _{\mathrm{c}}$ affect the interference, and we explore its potential applications, such as controllable field transmission and in situ switchable slow-fast light}. The $\rm S_{11}$ spectrum and group-time delay measurement are carried out on the VNA and then fitted by

\begin{equation}
T=\left\vert t_{\mathrm{p}}\right\vert
\end{equation}

\noindent and

\begin{equation}
\tau =-\frac{\partial\left[ \arg \left( t_{\mathrm{p}}\right) \right]}{\partial \Delta _{\mathrm{p}}},
\end{equation}

\noindent respectively.

\section{Phase induced interference and controllable microwave field transport}

\begin{figure}[hptb]
\includegraphics[width=0.6\linewidth]{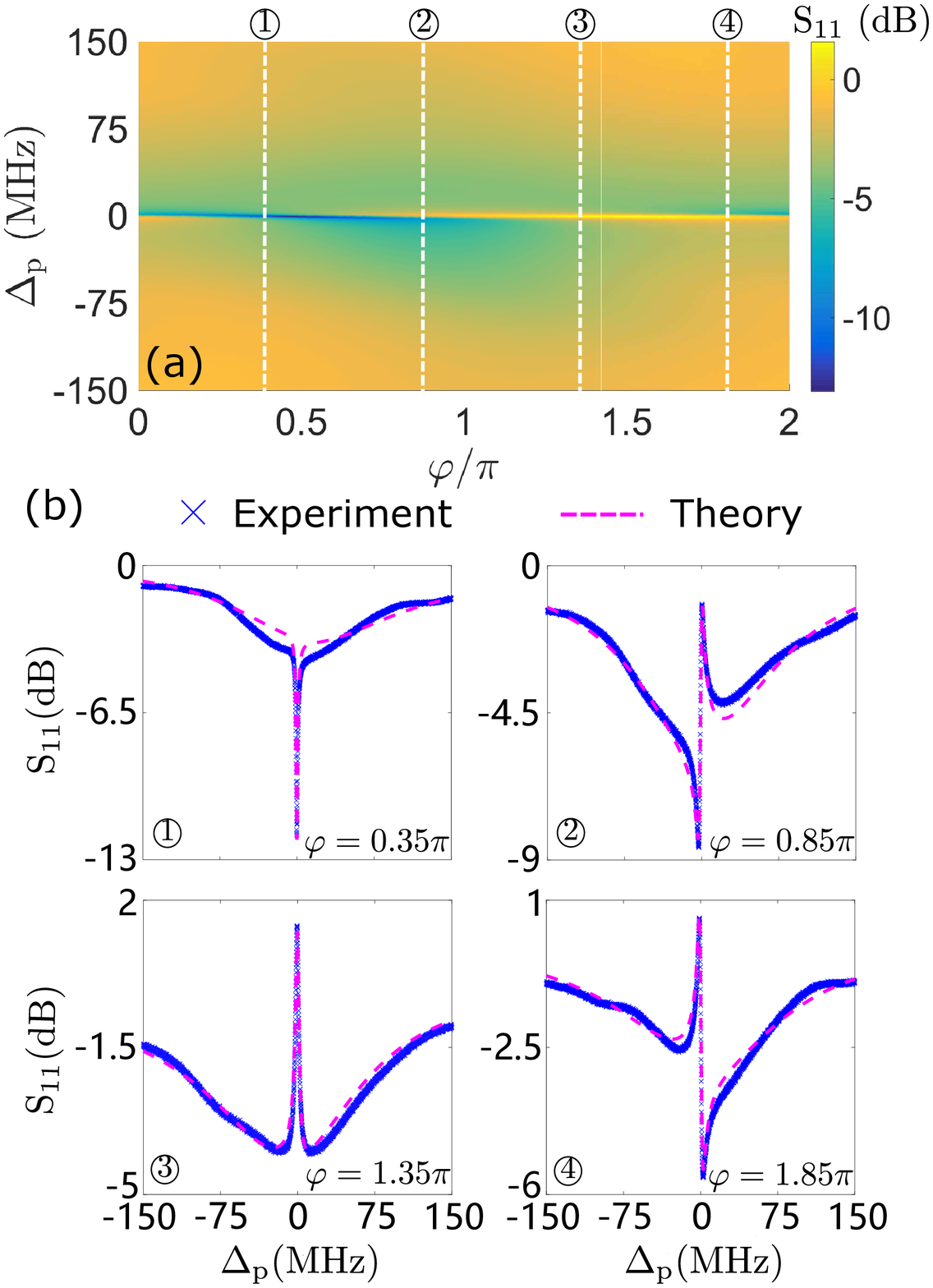}
\caption{\label{fig2} $\rm S_{11}$ spectrum versus relative phase difference $\varphi$. (a) Measured transmission spectrum $\rm S_{11}$ versus phase $\varphi$ and detuning $\Delta_{\mathrm{p}}$. The colors indicate the transmitted amplitudes in dB units. (b) Measured output spectrum $\rm S_{11}$ with phases: \textcircled{1} $\varphi=0.35\pi$, \textcircled{2} $\varphi=0.85\pi$, \textcircled{3} $\varphi=1.35\pi$, and \textcircled{4} $\varphi=1.85\pi$. Here, the pump-probe amplitude ratio is fixed at $\delta=1.7$. Red-solid lines are the corresponding theoretical results.}
\end{figure}

We first study how the phase of the magnon-pump tone affects the transmission of the cavity-probe field. In Fig.~\ref{fig2} (a), we present experimental results of the transmission, when the pump-probe ratio is $\delta=\varepsilon _{\mathrm{m}}/\varepsilon _{\mathrm{c}}=1.7$. In this setup, the phase $\varphi$ is continuously increased from 0 to 2$\pi$ using an I/Q mixer, and is shown in the $x$ axis of Fig.~\ref{fig2} (a). Then we conduct the $\rm S_{11}$ measurements and the recorded spectra are plotted versus the detuning frequencies $\Delta _{\mathrm{p}}$. The colors represent the relative steady-state output amplitude (in dB units) at different frequency and pump-probe ratios. Figure~\ref{fig2}(a) shows that the interference mainly happens around $\Delta _{\mathrm{p}}=0$ and can be controlled \textit{in situ} by changing the phase $\varphi$.

As shown in Fig.~\ref{fig2}(b), where $\varphi$ is set to $0.35\pi$, destructive interference happens and an obvious dip appears around $\Delta _{\mathrm{p}}=0$. This behavior can be regarded as MIABS. However, if we set $\varphi=1.35\pi$, constructive interference happens and an obvious amplification window appears around $\Delta _{\mathrm{p}}=0$. This behavior can be described as magnon-induced amplification (MIAMP). When $\varphi$ is set to $0.85\pi$ or $1.85\pi$, sharp and Fano-interference-like asymmetry spectra are observed even when the cavity and magnon are exactly resonant.

Although the interference originates from the coherent cavity-magnon coupling, Fig.~\ref{fig2} clearly shows that \textit{the phase $\varphi$ plays a key role in realizing an in situ tunable and controllable interference (e.g., constructive or destructive interference)}, which can be further engineered to control the probe-field transmission.
Note that in previous studies~\cite{zhang2014strongly} MIABS was only observed in the cavity overcoupling regime (i.e., $\eta _{\mathrm{a}}>1/2$) and MIT was only observed in the cavity undercoupling regime (i.e., $\eta _{\mathrm{a}}<1/2$). In contrast to this, here we realize a phase-dependent and switchable MIABS and MIT, \textit{as well as} MIAMP in a fixed undercoupling regime ($\eta _{\mathrm{c}}=0.19$ in our experiment). We emphasize that the destructive interference-induced MIABS is a unique result of phase modulation. The observed asymmetric Fano line shapes could be useful to realize Fano-interference sensors or precise measurements, using the magnon-pump method realized in our work.

\section{Amplitude ratio optimized magnon-induced-absorption}

\begin{figure}[hptb]
\includegraphics[width=0.5\linewidth]{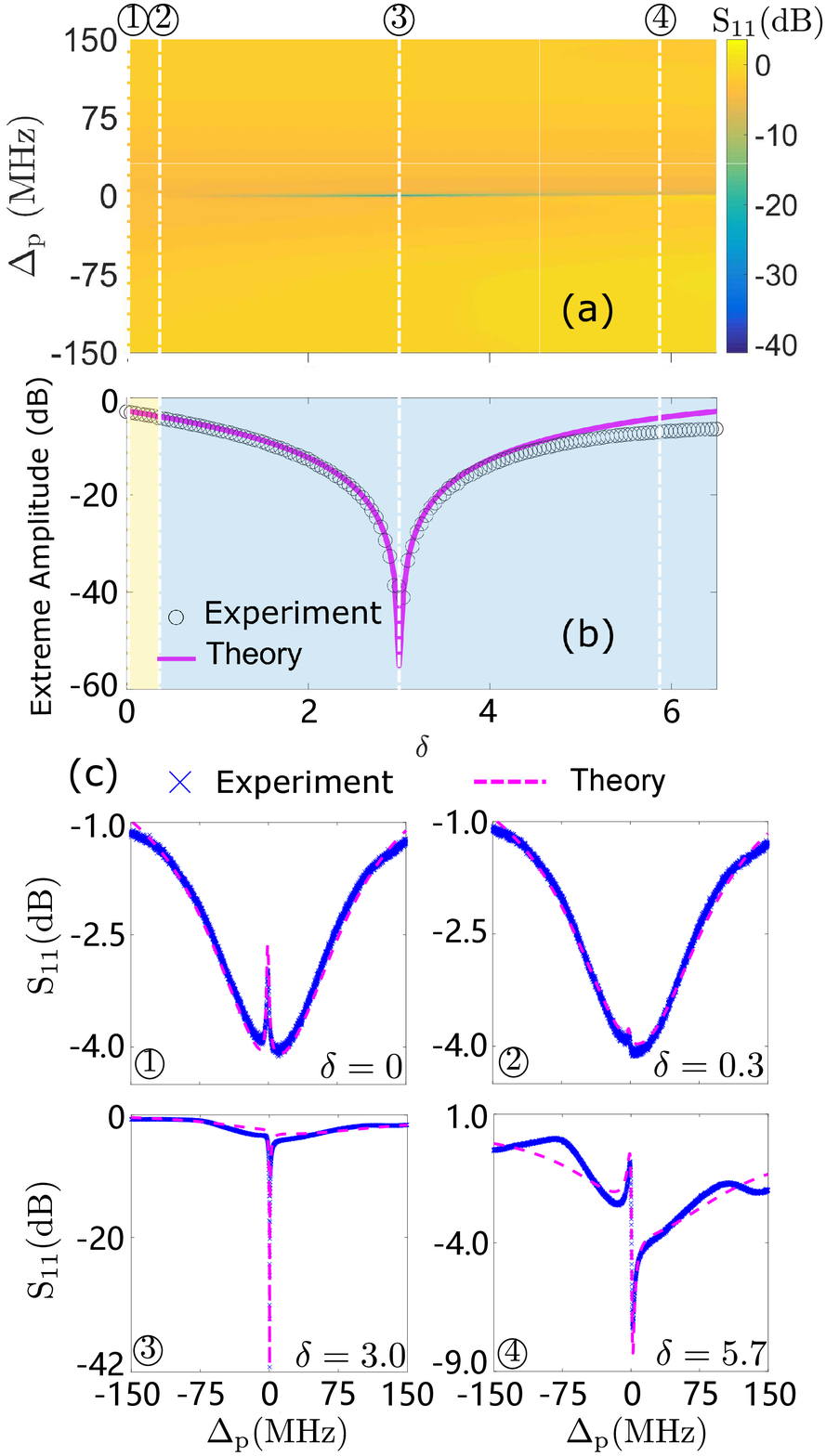}
\caption{\label{fig3} Measured transmission spectrum $\rm S_{11}$ versus pump-probe amplitude ratio $\delta$ with phase fixed at $\varphi=0.35\pi$. (a) Measured output spectrum versus amplitude ratio $\delta$ and detuning $\Delta_{\mathrm{p}}$. The colors indicate transmitted power in dBs. (b) The extreme values of the $\rm S_{11}$ transmission spectra of the output field versus the amplitude ratio parameter $\delta$. In the light-yellow (light-blue) regime, the extreme values represent the maximum (minimum) transmission amplitudes of the peaks (dips) around $\Delta_{\mathrm{p}}=0$. (c) Measured transmission spectrum $\rm S_{11}$ with amplitude ratio: \textcircled{1} $\delta=0$, \textcircled{2} $\delta=0.3$, \textcircled{3} $\delta=3.0$, and \textcircled{4} $\delta=5.7$. Red-solid lines are the corresponding theoretical results.}
\end{figure}

Recall the magnon-pump transmission coefficient $t_{\mathrm{pump}}$ in Eq.~(\ref{tpump}). There, the phase $\varphi$ determines the type of interference, e.g., constructive or destructive. However, the pump-probe ratio $\delta=\varepsilon _{\mathrm{m}}/\varepsilon _{\mathrm{c}}$ also affects the degree of interference, and thus can be used to control the probe-field transmissions $t_{\mathrm{p}}$. As shown in Fig.~\ref{fig3}(a), a color map is used to present the experiment results. Along the $x$ axis, the amplitude ratio $\delta$ is continuously increased from 0 to 6.5, by changing the overall voltage amplitude applied to the I and Q ports of an I-Q mixer. Then we conduct the $\rm S_{11}$ measurements and the steady-state output-field amplitudes are plotted versus the frequency detuning $\Delta_{\mathrm{p}}$. The colors in Fig.~\ref{fig3}(a) represent the relative strength of the steady-state output field (in dB units) at a different frequency. Here, the chosen phase $\varphi=0.35\pi$ results in MITs when $\delta<0.32$, while MIABSs dominate the output response in the regime $\delta>0.32$. We then study how the pump-probe ratio $\delta$ affects the central absorption window of the $\rm S_{11}$ spectra.

Figure~\ref{fig3}(a) shows that interference occurs around $\Delta _{\mathrm{p}}=0$ and is \textit{in situ} controlled by changing the pump-probe ratio $\delta$. The center blue-colored area represents an ideal absorption (transmission $T<0.01$) of the probe field.

Figure~\ref{fig3} (b) shows the extreme values of the transmission coefficients around $\Delta_{\mathrm{p}}=0$ versus the pump-probe ratio $\delta$. In the yellow area, we find the local maximum values of the MITs, and the local minimum values are found for MIABSs in the blue area. An obvious dip appears around $\delta=3$ and the minimum transmission value is less than 1\% (voltage amplitude ratio), which corresponds to an optimized and ideal probe-field absorption.

Figure~\ref{fig3}(c) shows the evolution process from MIT to MIABS by gradually increasing the pump-probe ratio $\delta$. When $\delta=0$, corresponding to case \textcircled{1} of Fig.~\ref{fig3}(c), our scheme recovers the traditional MIT case when no magnon pump is applied. When the magnon pump is introduced and its strength is continuously increased, the transparency window disappears and is replaced by an obvious absorption dip, as shown in cases \textcircled{2} and \textcircled{3} of Fig.~\ref{fig3}(c). With an even larger pump-probe ratio, the MIABS dips become asymmetry gradually, such as the spectrum in the case \textcircled{4} of Fig.~\ref{fig3}(c). Comparing with other results in Fig.~\ref{fig3}(c), we can find that the experimental data do no fit so well with the theory in case \textcircled{4} of Fig.~\ref{fig3}(c). This is induced by the additional cavity-antenna 2 coupling. Due to the existence of this tiny coupling, the magnon pump signal also pumps the cavity. With a modest magnon-pump strength, the additional cavity pump does not affect the system seriously, so that the theory fit the experiment data well. With a relatively strong magnon pump, the side effects of the additional cavity pump become larger, though it does not change the line shape. Therefore, the experiment data and theory do not fit so well when the magnon pump is relatively strong~\cite{suppl}. Similar phenomena can also be observed in the case \textcircled{4} of Fig.~\ref{fig4}(c).

We emphasize one main result of this paper: \textit{the absorption dips appear with an under-coupling coefficient of $\eta_{\textrm{a}}=0.19$ in our experiment. However, absorptions only happen in the overcoupling regime in traditional cases}. Moreover, Figs.~\ref{fig3}(a) and (c) show that \textit{$\delta$ can be used to switch the transmission behavior from the magnon-induced transparency to the magnon-induced absorption}. Note that the type of interference, destructive interference or constructive interference, depends on the value of the phase $\varphi$. However, the interference intensity is determined and optimized by the pump-probe ratio $\delta$. As shown in Fig.~\ref{fig3}(c), the dip of $\rm S_{11}$ is 42 dB lower than the baseline. The dip amplitude is quite close to zero, which indicates that a zero reflection is generated by the destructive interference.

\section{Amplitude ratio optimized magnon-induced-amplification}

\begin{figure}[hptb]
\includegraphics[width=0.5\linewidth]{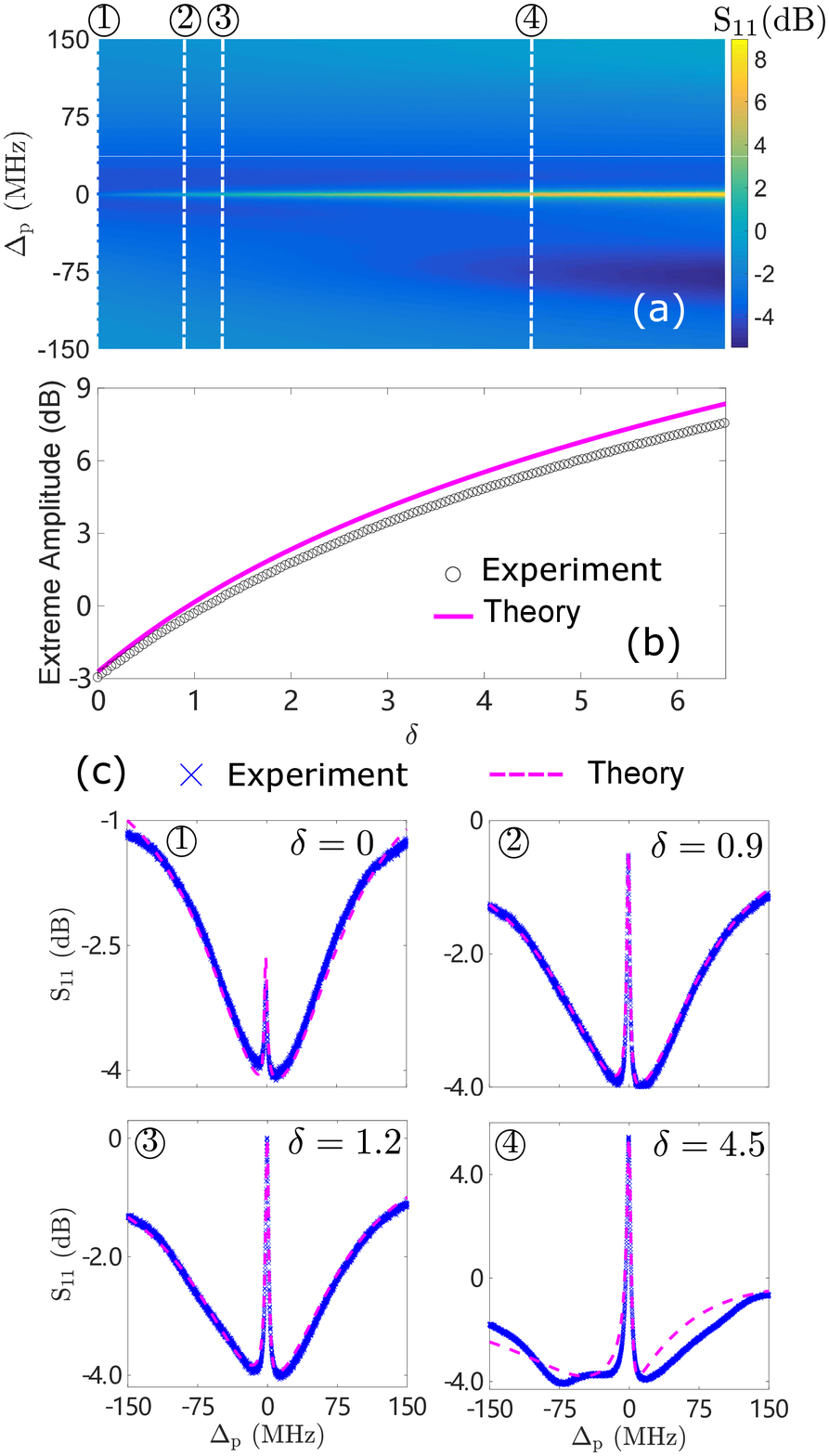}
\caption{\label{fig4} Measured transmission spectrum $\rm S_{11}$ versus pump-probe amplitude ratio $\delta=\varepsilon _{\mathrm{m}}/\varepsilon _{\mathrm{c}}$ with phase fixed at $\varphi=1.35\pi$. (a) Measured output spectra $\rm S_{11}$ versus amplitude ratio $\delta$ and frequency detuning $\Delta_{\mathrm{p}}$. The colors indicate the transmitted amplitude in dB units. (b) The extreme values of the $\rm S_{11}$ transmission spectra of the output field versus the amplitude-ratio parameter $\delta$. The extreme values represent the maximum transmission amplitude of the peaks around $\Delta_{\mathrm{p}}=0$. (c) Measured transmission spectra $\rm S_{11}$ with amplitude ratios: \textcircled{1} $\delta=0$, \textcircled{2} $\delta=0.9$, \textcircled{3} $\delta=1.2$, and \textcircled{4} $\delta=4.5$. The red-solid lines are the corresponding theoretical results.}
\end{figure}

We now study how the amplitude ratio of $\delta=\varepsilon _{\mathrm{m}}/\varepsilon _{\mathrm{c}}$ affects the MIAMP. In this case, the phase is fixed at $\varphi=1.35\pi$, where constructive interference dominates the transmission of the output field. As shown in Fig.~\ref{fig4}(a), a color map is used to present the measurement results. Along the $x$ axis, the pump-probe ratio $\delta$ is continuously increased from 0 to 6.5. Then we conduct the $\rm S_{11}$ measurement, and the steady-state transmission spectra are plotted versus the frequency detuning parameter $\Delta_{\mathrm{p}}$. The colors in Fig.~\ref{fig4}(a) represent the transmission amplitudes of the steady-state output field (in dB units) at different frequencies. We then study how the amplitude $\delta$ affects the center amplification window of the $\rm S_{11}$ spectra.

Figure~\ref{fig4}(a) clearly shows that constructive interference happens around $\Delta _{\mathrm{p}}=0$ and are \textit{in situ} controlled by changing the pump-probe ratio $\delta$. Magnon-pump-induced constructive interference happens when the probe field is nearly resonant with the cavity (also the magnon), and amplification windows appear. Around $\Delta_{\mathrm{p}}=0$, the color changes from light blue to orange when the pump-probe ratio $\delta$ increases from 0 to 6.5. This indicates that the higher amplification can be obtained with a larger pump-probe ratio $\delta$.

Figure~\ref{fig4}(b) shows how the peak values in the amplification window change versus the amplitude ratio $\delta$. The amplification coefficient is monotonously dependent on the increment of the pump-probe ratio $\delta$. Although the maximum pump-probe ratio is $\delta=6.5$ in our experiment, we emphasize that a higher transmission gain can be obtained using a larger pump power.

Figure~\ref{fig4}(c) clearly shows the evolution of the transmission spectrum from MIT to MIAMP when we gradually increase the pump-probe ratio $\delta$. When $\delta<1.2$, an obvious transparency window appears. When $\delta=1.2$, the peak value of the transparency window equals the value of the baseline, showing the ideal MIT phenomenon. Further increasing the pump strength, we can observe MIAMP. When $\delta=4.5$, an obvious amplification window appears, producing MIAMP.

Note that the phase is fixed at $\varphi=1.35\pi$ to produce constructive interference. When the amplitude ratio is set to $\delta=0$, i.e., no magnon pump, our scheme also recovers the traditional case without a magnon pump and only MIT is observed. This result is, of course, the same as case \textcircled{1} in Fig.~\ref{fig3}(c). We point out another main result that \textit{the pump-probe ratio $\delta$ can be used to realize and control the magnon-induced amplifications}. Figures~\ref{fig4}(a) and \ref{fig4}(c) show that $\delta$ can be used to switch the system response from MIT to MIAMP. Note that the interference type, such as constructive interference discussed here, depends on the value of the phase $\varphi$; however, the interference intensity is determined and optimized by the pump-probe ratio $\delta$.

\section{switchable fast- and slow-light based on the phase and amplitude ratio}

\begin{figure}[hptb]
\includegraphics[width=0.5\linewidth]{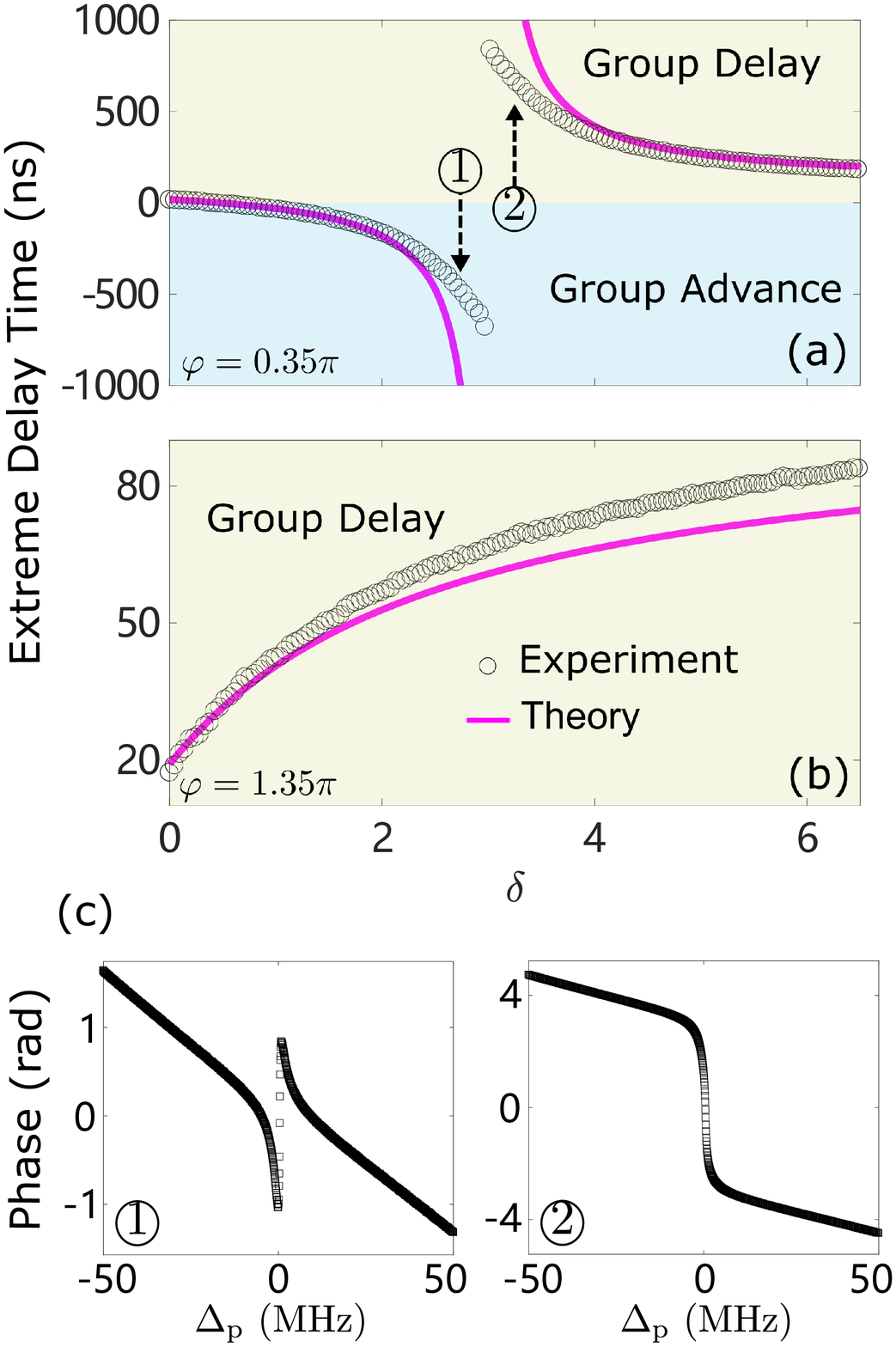}
\caption{\label{fig5} Measured time delay versus pump-probe ratio $\delta$ for the phase $\varphi=0.35\pi$ (a); and $\varphi=1.35\pi$ (b). Light-yellow area indicates the group-delay regime, and the light-blue area indicates the group-advance regime. (c) Measured unwrapped phase versus frequency detuning $\Delta_{\mathrm{p}}$ with $\delta=2.7$ [point $\textcircled{1}$ in (a)] and $\delta=3.3$ [point $\textcircled{2}$ in (a)] for $\varphi=0.35\pi$. }
\end{figure}

The group delay or advance of light always accompanies EIT or EIABS. In this experiment, we show that the group delay (slow light) and group advance (fast light) can also be realized in our cavity magnon-polariton system. Similar to the discussions above, the phase $\varphi$ is the key parameter that determines the interference type, e.g., destructive or constructive. Therefore, the phase $\varphi$ provides a tunable and \textit{in situ} switched group advance or delay of the probe field. The extreme values of the delay time are measured and presented in Fig.~\ref{fig5}, choosing the same phases $\varphi=0.35\pi$ and $\varphi=1.35\pi$, which are also used in Figs.~\ref{fig3} and \ref{fig4}, respectively.

In Fig.~\ref{fig5}(a), the phase is set to $\varphi=0.35\pi$. When we increase the pump-probe ratio $\delta$, a longer advance time is achieved, but immediately changes to time delay when $\delta>3.0$.  Further increasing $\delta$ reduces the delay time. In Fig.~\ref{fig5}(c), we present the phase of transmission signals at different probe frequencies with $\delta=2.7$ (case $\textcircled{1}$) and $\delta=3.3$ (case $\textcircled{2}$). The phase changes drastically around $\Delta_{\mathrm{p}}=0$ with opposite directions. The drastic changes of the phase result in a long advance or delay time, while the phase-change direction reversal results in the sharp transition from time advance to time delay. Accompanying the sharp transition in Fig.~\ref{fig5}(a), we observe the longest either delay or advance times. Therefore, the pump-probe ratio $\delta$ allows to \textit{optimize and switch the probe microwave from fast to slow light, or inversely}. Comparing the abrupt transition in Fig.~\ref{fig5}(a) with the zero reflection discussed in Sec. V, we find that the delay time abrupt transition and the zero reflection occur at the same parameter setup. It is notable that the discontinuity and abrupt transition are always accompanied by the zero reflection in coupled resonator systems. In Fig.~\ref{fig5}(b), we set the phase to $\varphi=1.35\pi$ and mainly observe constructive interference. In this case, the time delay monotonously increases with the pump-probe ratio $\delta$. Note that the pump-probe ratio used in Fig.~\ref{fig5}(b) is not its limitation, therefore longer delay times can be achieved by further increasing $\delta$.

Figure~\ref{fig5} also shows that when the amplitude ratio $\delta\leq3.0$, the delay time is a negative number which corresponds to fast light with $\varphi=0.35\pi$, and the positive delay time corresponds to slow light with $\varphi=1.35\pi$. Thus the phase parameter $\varphi$ can also be used to switch fast and slow light. When $\delta=0$, i.e., no magnon pump, our scheme recovers the traditional MIT and only a 16-ns delay time is achieved. By applying the magnon pump and optimizing $\varphi$ and $\delta$, \textit{the time delay, as well as advance, can be enhanced by nearly 2 orders of magnitude compared with the case without magnon pump}. For our scheme, the pump-probe amplitude ratio and phase difference mediated path interference can result in the zero reflection, which is accompanied with a delay time abrupt transition. In our experiment, Fig.~\ref{fig5}(a) clearly shows such an abrupt transition and greatly enhanced fast-slow light around this point. We can find that the experimental data deviates from the theoretical result around the abrupt transition. This is mainly induced by the imperfect system setups, such as limited output precision of AWG, imperfectness of the I-Q mixer and unstable magnon frequency~\cite{suppl}.

\begin{table}[]
\centering
\caption{\label{regime} Summary of MIT, MIABS, MIAMP and Fano resonance observed experimentally for different values in parameter space.}
\begin{tabular}{cl|c|c|c|c|c|c|}
\cline{3-8}
                                                       &                                & \multicolumn{6}{c|}{Amplitude Ratio $\delta$}                   \\ \cline{3-8}
                                                       &                                & 0 - 0.3 & 0.3  & 0.3 - 1.2 & 1.2           & 1.2 - 3.0 & $>3.0$ \\ \hline
\multicolumn{1}{|c|}{\multirow{2}{*}{Phase $\varphi$}} & \multicolumn{1}{c|}{$0.35\pi$} & MIT     & NULL & MIABS    & MIABS        & MIABS    & Fano   \\
\multicolumn{1}{|c|}{}                                 & \multicolumn{1}{c|}{$1.35\pi$} & MIT     & MIT  & MIT       & MIT (perfect) & MIAMP    & MIAMP \\ \hline
\end{tabular}
\end{table}

~\\

\vspace*{-6mm}
\section{Conclusion}
We experimentally study how the magnon pump affects the probe-field transmission, and the observed results are summarized in Table.~\ref{regime}. Two parameters, the relative phase $\varphi$ and the pump-probe ratio $\delta$ between pump and probe tones, are studied in detail. The main results of this work are as follows:

\begin{itemize}
\item the unconventional MIABS of the transmitted microwave field is observed with the cavity in the undercoupling condition;
\item MIAMP phenomena is realized in our experiment;
\item asymmetric Fano-resonance-like spectra are observed even when the cavity is resonant with the magnon;
\item by tuning the phase of the magnon pump, we can easily switch between MIT, MIABS and MIAMP;
\item by tuning the pump and probe ratio, the MIABS and MIAMP can be further optimized, accompanied by greatly enhanced advanced or slow light by nearly 2 orders of magnitude;
\item the tunable phase and amplitude ratio can lead to the zero reflection of the transmitted light and abrupt fast-slow light transitions.;
\item both the $\varphi$ and $\delta$ can be used to carry out the \textit{in situ} switch of fast and slow light.
\end{itemize}

Our results confirm that direct magnon pumping through the coupling loops provides a versatile route to achieve controllable signal transmission, storage, and communication, which can be further expanded to coherent state processing in the quantum regime. Furthermore, by exploiting multi-YIG spheres or multimagnon modes systems, the amplification or absorption bandwidth can be increased, resulting in a broadband coherent signal store device. The sharp peak and asymmetric Fano line shape indicate that our platform has great potential in the application of high-precision measurement of weak microwave fields~\cite{sedlacek2012microwave, liao2020microwave}. Our two-tone pump scheme and phase-tunable interference can also be accomplished in other coupled-resonator systems, such as optomechanical resonators, which explores effects of mechanical pump on light transmission~\cite{chang2011multistability, suzuki2015nonlinear, wu2018microwave, jia2015phase, xu2015controllable, jiang2019phase, lu2019selective, jing2015optomechanically, wang2014optomechanical}, and even in circuit-QED systems, in which photon transmission can be controlled through a circuit-QED system~\cite{liu2014blockade, gu2016polariton, wang2018two, sun2014electromagnetically}.

\begin{acknowledgments}
This work is supported by the National Key R\&D Program of China (Grant No. 2018YFA0306600), the CAS (Grants No. GJJSTD20170001 and No. QYZDY-SSW-SLH004), Anhui Initiative in Quantum Information Technologies (Grant No. AHY050000), and the Natural Science Foundation of China (NSFC) (Grant No. 12004044). F.N. is supported in part by: NTT Research, Army Research Office (ARO) (Grant No. W911NF-18-1-0358), Japan Science and Technology Agency (JST) (via the CREST Grant No. JPMJCR1676), Japan Society for the Promotion of Science (JSPS) (via the KAKENHI Grant No. JP20H00134 and the JSPS-RFBR Grant No. JPJSBP120194828), the Asian Office of Aerospace Research and Development (AOARD), and the Foundational Questions Institute Fund (FQXi) via Grant No. FQXi-IAF19-06.

Note added -- Recently, we become aware of a study presenting an infinite group delay and abrupt transition in a magnonic non-Hermitian system~\cite{yang2020unconventional}.

\end{acknowledgments}


%

\end{document}